# Strict Relationship: Potential - energy levels

F. Maiz[1] and M. Nasr[2]

[1]IPEIN, B.P:62, Merazka 8000, Nabeul, Tunisia.

[2]Faculty of mathematical sciences, university of Khartoum, p.o. box: 32, Sudan.

Actual Address:

[1]Department of Physics, College of Sciences, King Khalid University, Abha, Saudi Arabia. Email: fethi_maiz@yahoo.fr; fmaiz@kku.edu.sa

[2]Department of Mathematics, College of Sciences, King Khalid University, Abha, Saudi Arabia.  Email: mednasr@kku.edu.sa; m_nasr@yahoo.com

*Abstract:*

We have developed a new simple method to build the exact analytical expression of the eigenenergy as a function of the potential. The idea of our method is mainly based on the partitioning of the potential curve, solving the Schrödinger equation, realizing a discrete form of the energy quantification condition, and finally, deriving its integral form which permit to create a simple relation: Energy-potential. Our method has been applied to three examples: the well-known square well, the harmonic oscillators, and the Morse potential. Our non numerical method is more realistic, simpler, with high degree of accuracy, both satisfactory and not computationally complicated, and applicable for any forms of potential. Our results agree very well with the preceding ones.

*1. Introduction*

*One of the challenging problems in nonrelativistic quantum mechanics is to find exact solutions to the Schrödinger equation for potentials that proves to be useful to model phenomena in nuclear physics, solid state physics, molecular-atomic physics, and laser theory. Since its appearance, there have been continual researches for solving the Schrödinger equation with any potential by using different methods such as the factorization [1], the path integral [2], the group theoretical (algebraic method) [3], the 1/N expansion [4], the analytic continuation [5], the eigenvalue moment method [6], the power series expansion [7] and the supersymmetric quantum mechanics [8] could be enumerated amongst other methods of the solutions of the wave equations. Laplace transforms, which are integral transforms, are comprehensively useful in physics and engineering [9]. Anharmonic oscillators are themselves interesting since the real world deviates from an id0ealized picture of harmonic oscillators because of self interactions and interactions between them. Therefore, anharmonic systems have been studied extensively both analytical and numerical. The spectra of many important potential functions frequently encountered in quantum mechanics cannot be obtained exactly. Moreover, in most cases, the conventional approximate methods commonly discussed in most standard textbooks [10], are either unsatisfactory or computationally complicated.*

*In this paper we propose a very simple new method which allows the determination of the energy levels for any forms of potential. This method is with high degree of accuracy and is both satisfactory and not computationally complicated. This paper is organized as follows. In Sect. 2, we present our new approach which is based on the partitioning of the potential curve into n+1 small intervals, solving the Schrödinger equation in each subintervals, and writing the continuity conditions of each solutions at each subintervals boundary, which leads to the energy quantification condition, so to the energy levels, and*

*finally, we explain how obtain the accurate energy levels. In Sect.3.a, we show that the energy quantification condition can be simplified, and manage its simple integral form. As application, we study, in Sect.3.b, the of infinite square well case, in Sect.3.c, the well known harmonic oscillator, and the Morse potential played, in Sect.3.d. The calculated eigenvalues are on high accuracy and agree very well with the preceding ones. The corresponding wavefunctions may be deduced straightforwardly by solving the Schrödinger equation for each energy level.*

## 2. Formulation

*To calculate the eigenvalues of energy of harmonic and anharmonic oscillators bounded by infinity high potentials* $V = \infty$ *at* $x = a$ *( point A ) and* $x = b$ *( point B ), here* $b - a = 2L$ *the large finite values of the separation of the walls ( see Fig.1.), it is necessary to solve the one-dimensional Schrödinger equation ( throughout this paper, we assume that* $\hbar = 1$ *and* $2m = 1$ *):*

$$\frac{d^2\Psi(x)}{dx^2} + (E - V(x))\Psi(x) = 0 \qquad (1)$$

*As described by Fig. 1, the potential energy* $V(x)$ *is a continuous function (solid line) between* $x = a$ *and* $x = b$*, in order to determinate the energy levels, we start by partitioning the interval [a,b] into n+1 small subintervals* $I_i = [x_i, x_{i+1}]$ *each with width* $h$*, where* $h = (b-a)/(n+1)$ *and* $x_i = a + ih$ *for* $i = 0,...,n$.

*The midpoint of each of these subintervals is given by* $\rho_i = (x_{i+1} - x_i)/2$*. We evaluate our function* $V(x)$ *at the midpoint of any subinterval, and prolong it by adding two large intervals* $I_{-1}$ *and* $I_{n+1}$ *i.e.:*

$$\begin{cases} \text{for } x \leq a, & x \in I_{-1} : & V(x) = \infty \\ \text{for } x_i \leq x \leq x_{i+1}, & x \in I_i : & V(x) = V(\rho_i) \\ \text{for } b \leq x, & x \in I_{n+1} : & V(x) = \infty \end{cases} \qquad (2)$$

Let $k_i^2 = (V_i - E)$, *writing and solving equation (1) in each subinterval leads to the following solutions:*

$$\Psi_i(x) = X_i \exp(k_i x) + Y_i \exp(-k_i x) \qquad (3)$$

*Where $X_i$ and $Y_i$ are constants. For the two intervals $I_{-1}$ and $I_{n+1}$, the potential is infinity, and $\Psi(a) = \Psi(b) = 0$. The functions $\Psi_i(x)$ are twice derivable with respect to x, and they are continuous. The continuity of the solutions $\Psi_i(x)$ and their derivates at the different points $x_i$ allows to the elimination of $X_i$ and $Y_i$, and leads to the following equation known as the energy quantification condition:*

$$B_n(E) = a_n P_n + b_n Q_n = 0 \qquad (4)$$

*here:* $c_{ij} = c_{i,j} = \sqrt{(V_i - E)/(V_j - E)}$, $a_i = \exp(-k_i h)$, $b_i = \exp(k_i h)$, $Q_0 = -1$, $P_0 = 1$,

$P_i = (c_{i,i-1} + 1) a_{i-1} P_{i-1} + (c_{i,i-1} - 1) b_{i-1} Q_{i-1}$, *and* $Q_i = (c_{i,i-1} - 1) a_{i-1} P_{i-1} + (c_{i,i-1} + 1) b_{i-1} Q_{i-1}$

*One can proof this expression in the general case by the recurrence method as described by Maiz [11, 12]. The energy levels are obtained by the energy values for which the curve of $B_n(E)$ meets the energy axis. We consider the function F(E) defined as: $F(E) = \text{signum}(B_n(E))$, where the signum function computes the sign of the leading coefficient of the expression [if $x \neq 0$ then $\text{signum}(x) = x / \text{abs}(x)$, and $\text{signum}(0) = 0$]. The energy levels are indicated by a vertical segments perpendicular to the energy axis which constitute the curve of the function F(E). In this case, the energy levels are determinated with a great precision. The energy quantification condition may be simplified according to the potential function, for example, the term $c_{ij}$ values are limited.*

### 3. Energy Quantification Condition

#### 3.a. Simplification

*In order to obtain the exact energy levels, we must use great values for the integer number n. In this case* $V_i$ *is in close proximity to* $V_{i-1}$ *and the term* $c_{i,i-1} = \sqrt{(V_i - E)/(V_{i-1} - E)}$ *can take only three values:*

    *1- if* $V_i$ *and* $V_{i-1}$ *are greater then E :* $c_{i,i-1} = 1$.

    *2- if* $V_i$ *and* $V_{i-1}$ *are smaller then E :* $c_{i,i-1} = 1$.

    *3- if* $V_i$ *is greater then E and* $V_{i-1}$ *is smaller then E, and conversely, then:*

    $c_{i,i-1} = +I$ *or* $-I$. *Here, I is the complex number which satisfies:* $I^2 = -1$, *and if* $c_{i,i-1} = +I$ *then* $c_{i-1,i} = -I$, *and conversely, because their product is equal to 1* ( $c_{i,i-1}c_{i-1,i} = 1$ ). *By convention we take the value* $+I$ *for* $c_{i,i-1}$ *if the potential function increases, and the value* $-I$ *in the opposite case, and* 1 *when the potential function is constant.*

    *These simplifications of* $B_n(E)$, *which depends mainly on the potential expression, generate a very simple expression for* $B_n(E)$, *in fact when* $c_{i,i-1} = 1$ *the second term of* $P_i$ *and the first term of* $Q_i$ *vanish. Sect.3.b. explains, in the case of the infinite square well, the simplifications of the energy quantification condition, in its discrete form. The search of the unknown great integer number n may takes much time and always gives doubtful results. So, it is preferable to use an integral form. For great values of n the width* $h = (b-a)/(n+1)$ *is very small and may be replaced by dx, the terms* $a_i = \exp(-k_i h)$ *and* $b_i = \exp(k_i h)$ *become* $a_i = \exp(-k_i dx)$ *and* $b_i = \exp(-k_i dx)$, *and the energy quantification condition:* $B_n(E) = a_n P_n + b_n Q_n = 0$ *becomes for the potential described by Fig.1:*

$$B(E) = \frac{I_{2p} + I_{2m}}{I_{2m} - I_{2p}} - I \frac{I_{1p}^2 + I_{3p}^2}{(I_{1p} I_{3p})^2 - 1} = 0,$$

*Here:* $k(x) = \sqrt{V(x) - E}$, $I_{2p} = 1/I_{2m} = \exp(\int_{x_C}^{x_D} k(x)dx)$, $I_{1p} = \exp(\int_{x_A}^{x_C} k(x)dx)$, *and* $I_{3p} = \exp(\int_{x_C}^{x_D} k(x)dx)$.

*In Sect.3.c. we use integral form of the energy quantification condition to study the harmonic oscillator's case.*

### 3.b. The infinite square well

*In the case of infinite square well ( $V(x) = 0$ for $|x| \leq L$ and $V(x) = \infty$ for $|x| > L$ ), for all values of i: $E > V_i$, $c_{i,j} = 1$, $k_i = \sqrt{-E} = I\sqrt{E}$, and the energy quantification condition $B_n(E) = a_n P_n + b_n Q_n = 0$ becomes: $B_n(E) = \prod_{i=0}^{i=n} a_i - \prod_{i=0}^{i=n} b_i = 0$, this expression may be simplified to:*

$B_n(E) = \exp(-\sum_{i=0}^{i=n} Ih\sqrt{E}) - \exp(\sum_{i=0}^{i=n} Ih\sqrt{E}) = \exp(-I(n+1)h\sqrt{E}) - \exp(I(n+1)h\sqrt{E}) = 0$, *and finally, we obtain the well known condition:* $\sin(2L\sqrt{E}) = 0 \Rightarrow E = \frac{p^2 \pi^2}{(2L)^2}$, *here p is a non null integer number, and 2L the well large. We notice that the energy levels increase with the integer number p, and decrease with the separation value of the walls. The application of the energy quantification condition, in its integral form, gives the same results, in fact: first, if the segment CD represent the unknown energy value which we find, we remark that the points A and C are confused and also for the points B and D. Second $I_{1m} = I_{1p} = I_{3m} = I_{3p} = 1$, and $I_{2p} = 1/I_{2m} = \exp(I2L\sqrt{E})$. Third, the condition: $B(E) = 0$ leads to $I_{2m} - I_{2p} = 0$, and $\exp(I2L\sqrt{E}) - \exp(-I2L\sqrt{E}) = 0$, so the result $\sin(2L\sqrt{E}) = 0$ is proved.*

### 3.c. The harmonic oscillator

*As application we choose to study the well known harmonic oscillator. The quantum harmonic oscillator is the quantum mechanical analogue of the classical harmonic oscillator. It is one of the most important model systems in quantum mechanics because an arbitrary potential can be approximated as a harmonic potential at the vicinity of a stable*

*equilibrium point. Furthermore, it is one of the few quantum mechanical systems for which a simple exact solution is known. In the one-dimensional harmonic oscillator problem, a particle is subject to a symmetric potential* $V(x) = x^2$. *In this case:* $x_C = -\sqrt{E}$, $x_D = \sqrt{E}$, $I_{1m} = I_{3m}$, *and* $I_{1p} = I_{3p} = \exp(L(\sqrt{L^2 - E}) - E\ln(L + \sqrt{L^2 - E}) + E\ln(\sqrt{E}))/2$, *and* $I_{2p} = \exp(I\pi E/2)$.

*One can easily found that the energy quantification condition* $B(E) = 0$ *may be written as* $\cot(\pi E/2) = [\sinh(L(\sqrt{L^2 - E}) - E\ln(L + \sqrt{L^2 - E}) + E\ln(\sqrt{E}))]^{-1}$. *For the real harmonic oscillator, as* $L \to \infty$ *then* $\sinh(L(\sqrt{L^2 - E}) - E\ln(L + \sqrt{L^2 - E}) + E\ln(\sqrt{E})) \to \infty$, *and finally:* $\cot(\pi E/2) = 0$, *which leads to the discrete values of the energy levels already known for* $p = 0,1,2,3....$ *[13]:* $E_p = 2(p + \frac{1}{2})$. *This energy spectrum is remarkable for three reasons. Firstly, the energies are "quantized", and may only take the discrete values 1, 3, 5, and so forth. This is a feature of many quantum mechanical systems. Secondly, the lowest possible energy is not zero, but 1, which is the ground state energy. The final reason is that the energy levels are equally spaced, unlike the infinite square well case.*

### 3.d. The Morse potential

*The Morse potential played an important role in many different fields of the physics such as molecular physics, solid state physics, and chemical physics, etc. This potential has been studied by many different approaches such as the standard confluent hypergeometric functions [14], the algebraic method [15], the supersymmetric method [16], the coherent states [17], the controllability [18], the series solutions of Morse potential with the mass distribution [19], etc. G. Chen [20] had obtained the exact bound state solutions of the one-dimensional Schrödinger equation with the Morse potential by the way of Laplace transforms. S.-H.Dong and G.-H.Sun [21] have carried out the analytical solutions of the*

*Schrödinger equation by the series expansion method. H. Sun [22] have analytically derived the eigenenergies of the Morse potential by the analytical transfer matrix method.*

*In this paper, the Morse potential has the following form:* $V(x) = V_0(1 - \exp(-\lambda x))^2$, *where* $V_0$ *and* $\lambda$ *are the depth and range parameters, respectively, the exact energy eigenvalues are given by:* $E_n = 2\lambda\sqrt{V_0}\left[(n+\frac{1}{2}) - (n+\frac{1}{2})^2 \frac{\lambda}{2\sqrt{V_0}}\right]$.

*These energies levels were obtained exactly by Alhendi [23], for a bounded Morse potential confined to the interval* $-2 \leq x \leq 2$. *In this case* $x_C = \frac{-1}{\lambda}\ln(1+\sqrt{E/V_0})$, $x_D = \frac{-1}{\lambda}\ln(1-\sqrt{E/V_0})$, , *and* $I_{2p} = \exp(\frac{I\pi}{\lambda}(\sqrt{V_0} \pm \sqrt{V_0 - E}))$, *and the energy quantification condition* $B(E) = 0$ *leads to* $\cos(\frac{\pi}{\lambda}(\sqrt{V_0} \pm \sqrt{V_0 - E})) = 0$, *and the energy values* $E_n = 2\lambda\sqrt{V_0}\left[(n+\frac{1}{2}) - (n+\frac{1}{2})^2 \frac{\lambda}{2\sqrt{V_0}}\right]$.

*4. Conclusion*

*In this paper, we have developed a new simple method to build the exact analytical expression of the eigenenergy as a function of the potential. The idea of our method is mainly based on the partitioning of the potential curve into (n+1) small intervals, solving the Schrödinger equation in each subintervals, and writing the continuity conditions of each solutions at each subintervals boundary, realizing a discrete form of the energy quantification condition, and finally, deriving its integral form which let to create a simple relation: Energy-potential. Our method has been applied to three examples: the well-known square well, the harmonic oscillators, and the Morse potential. Our non numerical method is more realistic, simpler, with high degree of accuracy, both satisfactory and not computationally complicated, and applicable for any forms of potential. Our results agree very well with the preceding ones.*

*Figures:*

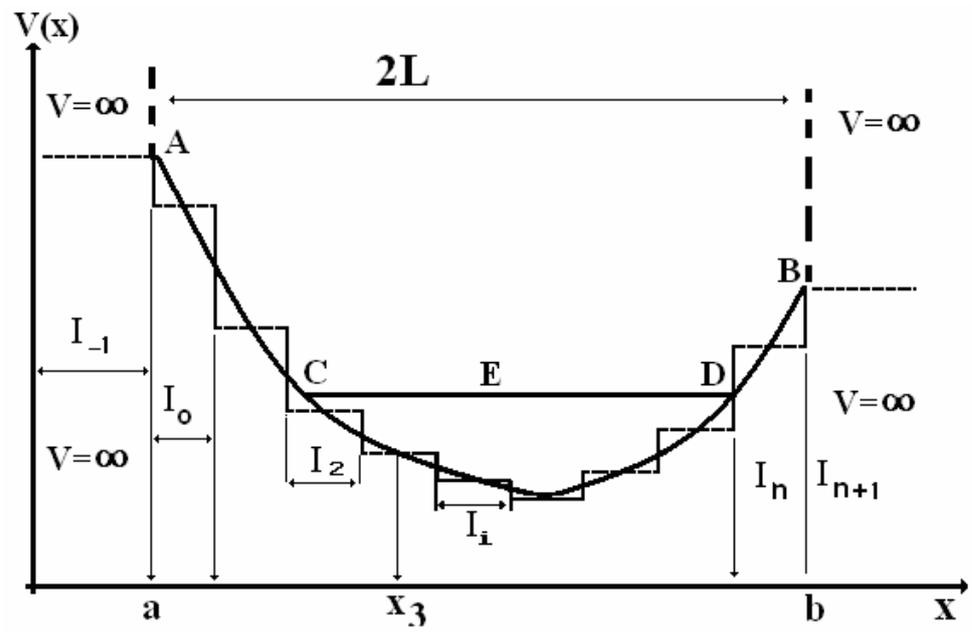

*Fig. 1: repartitioning of the potential curve into (n+1) small intervals.*